\documentclass{article}

\usepackage{arxiv}

\usepackage{url,hyperref,lineno,microtype}
\usepackage[onehalfspacing]{setspace}

\usepackage{amsmath}
\usepackage{dsfont}
\usepackage{verbatim}
\usepackage{bm}
\usepackage{caption}
\usepackage{graphicx}
\usepackage{comment}

\usepackage{natbib}
\usepackage{subcaption}

\usepackage{longtable}
\usepackage{rotating}
\usepackage{array}
\usepackage{multicol}
\usepackage{multirow}
\usepackage{xcolor,colortbl}

\newcommand{\review}[1]{\textcolor{black}{#1}}

\author{
  Hugo Breuillard \\
Laboratoire de physique des plasmas (LPP)\\
Palaiseau, France \\
 \texttt{hbreuill@gmail.com} \\
   \AND
 Romain Dupuis \\
 Center for Mathematical Plasma Astrophysics (CmPA) \\
 KU Leuven, Leuven, Belgium \\
  \AND
  Alessandro Retino \\
  Laboratoire de physique des plasmas (LPP)\\
Palaiseau, France \\
  \AND
  Oliver Le Contel \\
  Laboratoire de physique des plasmas (LPP)\\
Palaiseau, France \\
  \AND
  Jorge Amaya \\
   Center for Mathematical Plasma Astrophysics (CmPA) \\
 KU Leuven, Leuven, Belgium \\
  \AND
  Giovanni Lapenta \\
   Center for Mathematical Plasma Astrophysics (CmPA) \\
 KU Leuven, Leuven, Belgium \\
}

\begin{document}

\title{Automatic classification of plasma regions in near-Earth space with supervised machine learning: application to Magnetospheric Multi Scale 2016-2019 \review{observations}}

\maketitle

\begin{abstract}
The proper classification of plasma regions in near-Earth space is crucial to perform unambiguous statistical studies of fundamental plasma processes such as shocks, magnetic reconnection, waves and turbulence, jets and their combinations. The majority of available studies have been performed by using human-driven methods, such as visual data selection or the application of predefined thresholds to different observable plasma quantities. While human-driven methods have allowed performing many statistical studies, these methods are often time-consuming and can introduce important biases. On the other hand, the recent availability of large, high-quality spacecraft databases, together with major advances in machine-learning algorithms, can now allow meaningful applications of machine learning to \review{in-situ} plasma data. In this study, we apply the fully convolutional neural network (FCN) deep machine-leaning algorithm to the recent Magnetospheric Multi Scale (MMS) mission data in order to classify \review{ten} key plasma regions in near-Earth space for the period 2016-2019. For this purpose, we use available intervals of time series for each such plasma region, which were labeled by using human-driven selective downlink applied to MMS burst data. We discuss several quantitative parameters to assess the accuracy of both methods. Our results indicate that \review{the FCN method is reliable to accurately classify labeled time series data} 
 since it takes into account the dynamical features of \review{the plasma data in} each region. We also present good accuracy of the FCN method when applied to unlabeled MMS data. Finally, we show how this method used on MMS data can be extended to data from the Cluster mission, indicating that such method can be successfully applied to any in situ spacecraft plasma database.
\end{abstract}


\section{Introduction}
Near-Earth space regions classification is a very challenging topic in space physics because of the strong plasma dynamics resulting from the interaction between the propagating solar wind and the standing Earth's magnetosphere. This obstacle forms a collisionless bow shock~\citep{kivelson1995introduction} which can reflect and backpropagate energetic ions into the supersonic solar wind, forming the ion foreshock region upstream. However, most of the plasma is decelerated and heated downstream to a subsonic flow which forms a sheath around the Earth's magnetosphere called the magnetosheath. The boundary on which magnetosheath and magnetospheric pressures balance is called the magnetopause. The plasma flows along the magnetopause towards the magnetotail\review{, driving the frozen-in magnetic field lines through the lobes, plasma sheet boundary layer and ultimately the plasma sheet, in which they eventually reconnect~\citep{1963Dungey}. This large-scale solar-terrestrial interaction process is dominated by the highly-variable solar wind conditions, but also by localized and intermittent small-scale processes in each region.} As a result, the plasma throughout the near-Earth space is highly dynamic which make it impossible to classify the different regions using only spacecraft location along its orbit.

Yet, this classification is needed for statistical studies of the plasma properties of the different regions. These large-scale statistics are allowed by the ever-growing quantity of available in-situ data from space missions that cover most of the near-Earth space. A large amount of available data has been used to identify near-Earth space regions using human-driven (i.e., threshold-based) methods, e.g.,~\citep{jelinek2012}. \review{However, finding the optimal set of thresholds which characterizes every region of the magnetosphere is time-consuming and presumably not flexible enough with regard to the large variability expected from the strong plasma dynamics. This may introduce important biases depending on the chosen thresholds, and may also result in many "unclassified" regions (i.e., a significant loss of information).}









\review{One way to bypass this issue} and benefit from this large amount of data is to make use of supervised learning, the most common form of machine learning. The role of the supervised algorithms is to learn the relationship between data instances and an associated label for each data instance. For \review{example}, the text from an email may represent the data instance and the label is a binary encoding the presence or not of spam in order to detect and filter spams from emails.
Supervised methods aim at expanding human knowledge automatically by identifying the intrinsic differences between labeled points in a dataset. This knowledge is then generalized to other unlabeled datasets. Such machine learning techniques show extremely promising (often state-of-the-art) results in various tasks, such as image and speech recognition, analyzing particle accelerator data, or natural language understanding~\citep{lecun2015deep}.
They have already been used specifically in heliophysics: to classify time series into several categories defined by the solar origin of the wind~\citep{camporeale2017classification}, to detect magnetospheric ultra-low frequency waves~\citep{balasis2019machine}, to forecast solar flares~\citep{nishizuka2017solar}, or in particular to identify different regions of the near-Earth space~\citep{olshevsky2019, nguyen2019}. Nevertheless, these models remain usually region- or mission-specific because the process of human-labeling is a time-consuming and tedious task and there is still a blatant lack of human labels to cover the whole near-Earth space.

However, the Magnetospheric Multi Scale mission (MMS)~\citep{burch2016magnetospheric}, launched in 2015 and whose orbit covers most of near-Earth space, has enabled such a dataset to be built. This mission uses the so-called Scientist In The Loop (SITL) system (see the following section for more details) to human-label phenomena and regions of interest. The SITL process involves that an expert scientist is designated to select data of interest that will be transmitted to the ground as high-time resolution (burst) data. The selected intervals are tagged with timestamps and a comment, which are reported in the SITL reports. This system has been active since July 2015 and thus provides more than 4 years of commented data as of now. This makes it the biggest dataset of commented events and regions to our knowledge. We developed an automatic parser to convert all the comments, written by different scientists, in a unified list of labels that can be used by a machine learning algorithm. 

All SITL comments are associated with a time interval, therefore the temporal variations of the measurements can be used to predict the label of the signals. Thus, each data instance can be interpreted as a time series and the near-Earth classification using SITL data can be seen as a time serie\review{s} classification (TSC), a very challenging problem in data mining and machine learning~\citep{esling2012time, fawaz2019deep}. The main difference with classical classification problems comes from the ordering of the data. The important features helping to discriminate the labels are mainly found in the ordering of the values while classical methods only consider the values at a given time. We employ two different techniques to solve this multivariate and multicategory classification task: the multilayer perceptron is applied naively on instantaneous data and is considered as a baseline for comparison and the convolutional neural network is applied on time series. 

In this article, we use the magnetic field and particle measurements of the MMS mission to establish a vast and reproducible automatic detection of 8 plasma regions that cover almost the entire near-Earth space. First, we introduce the MMS and Cluster missions, and how we used the SITL reports to build a large labeled dataset of magnetic field, plasma moments and plasma distributions from the 2016-2019 period. Then, we present the two machine learning algorithms we used and their characteristics, as well as the metrics to measure their performances. Later, we present their respective results for the MMS mission and the adaptation to the Cluster mission. Finally, we discuss these results and show our conclusions.

\section{Data}
A fundamental and time-consuming step in any machine learning application is collecting, cleaning, and labelling data. This is a very important task as the quality of the data defines accuracy and the generalisation capability of the model.

\subsection{The MMS and Cluster missions}
MMS is a NASA space mission, launched in 2015, designed to study the electron-scale physics in Earth's magnetosphere and in particular where magnetic reconnection occurs~\citep{burch2016magnetospheric}. Its \review{equatorial} orbit is optimized to spend extended periods in locations where reconnection is known to occur and thus covers the majority of the near-Earth space key regions. The mission is composed of four identical spacecraft flying in an adjustable tetrahedral formation and its highly-elliptic orbit also covers almost all regions of near-Earth space. The direct-current (DC) magnetic field data are provided by the Fluxgate Magnetometer (FGM) from the FIELDS instrument suite~\citep{torbert2016fields, russell2016magnetospheric} with a temporal resolution of 0.1s in "survey" mode and the plasma parameters by the Fast Plasma Investigation (FPI) instrument~\cite{pollock2016fast} with a temporal resolution of 4.5s in "fast" mode.

Cluster is an ESA space mission, launched in 2000, whose aim is to study the ion-scale physics of Earth's magnetic environment and its interaction with the solar wind. The mission is also composed of four identical spacecraft flying in a tetrahedral formation and its highly-elliptic \review{polar} orbit also covers almost all regions of near-Earth space. The magnetic field data are provided by the Fluxgate Magnetometer with a temporal resolution of 4s~\citep{2001Balogh} and the plasma parameters by the Hot Ion Analyzer instrument~\citep{2001Reme} when the instrument was working under the magnetosphere or the magnetosheath mode.

\subsection{The labels datasets}
One of the most important innovation brought by the MMS mission is its burst data management and selection system. The MMS spacecraft collect a combined volume of $\sim100$ gigabits per day of particle and field data. On average, only 4 gigabits of that volume can be transmitted to the ground. With nested automation and "Scientist-in-the-Loop" (SITL) processes, this system is designed to maximize the value of the burst data by prioritizing the data segments selected for transmission to the ground.

Concretely, the SITL system consists of a manual selection process by a scientific expert designated to eyeball daily survey (low-time resolution) data and pick time intervals of interest that will be transmitted to the ground as burst (high-time resolution) data. The selected intervals are tagged with timestamps and a comment, which usually includes the type of event selected and eventually the near-Earth region in which it occurred. For each orbit, these selections are written up in a "SITL report", in the form of a text file notably. \review{An example of these reports can be found in~\citet{argall2020mms}.}

We developed a Python code which parses these text files to extract the timestamps and comments by identifying keywords associated with the near-Earth region where the event occurred. Each time interval was then associated with a label indicating where the spacecraft was located as follows:

\begin{table}[h]
	\centering
	\begin{tabular}{c|c|}
		\cline{1-2}
		\multicolumn{1}{|l|}{Regions}        & Labels \\ \hline
		\multicolumn{1}{|l|}{Solar wind}   & SW      \\ 
		\multicolumn{1}{|l|}{Ion foreshock}  & FS      \\ 
		\multicolumn{1}{|l|}{Bow shock}    & BS     \\ 
		\multicolumn{1}{|l|}{Magnetosheath}      & MSH      \\ 
		\multicolumn{1}{|l|}{Magnetopause}     & MP      \\ 
		\multicolumn{1}{|l|}{Boundary layer}      & BL      \\ 
		\multicolumn{1}{|l|}{Magnetosphere}  & MSP      \\
		\multicolumn{1}{|l|}{Plasma sheet}  & PS      \\
		\multicolumn{1}{|l|}{Plasma sheet boundary layer}  & PSBL      \\
		\multicolumn{1}{|l|}{Lobe}  & LOBE      \\ \hline
	\end{tabular}
	\caption{Labels for the 10 different near-Earth regions in our model.}
	\label{tab:region_labels}
\end{table}

If a region cannot be identified from the SITL comment, then the time interval is rejected. If FGM or FPI data is not available during a labeled time interval, then the time interval is also rejected. Using this technique, from the date at which the SITL reports are available as text files, i.e., April 2016, to the end of 2019, \review{we collected $7,832$ labeled time intervals relevant for our study}. We note here that the number of occurrences for each regions were somewhat unbalanced, so we added a total of \review{$605$} time intervals labeled by hand to undersampled regions \review{(examples of typical plasma parameters for the different magnetospheric regions can be found in textbooks, e.g.~\citet{baumjohann1996book})}, bringing the total to \review{$8,437$} labeled time intervals of various lengths. \review{We resampled the label data to the same cadence as FPI data, i.e., 4.5s, representing a total of $1,331,133$ labeled data points}. This constitutes the biggest dataset of labeled time intervals for near-Earth space to our knowledge.

Regarding the Cluster mission labels, we use the dataset presented in~\citet{nguyen2019}. This dataset covers three near-Earth space regions, namely the magnetosphere, the magnetosheath and the solar wind, over a 2-year time period (2005-2006). The data is resampled to a 1 minute resolution, yielding a total of $148,762$ labelled points.

\subsection{Final data set}
Once the labels are defined, we build the final dataset that will feed the ML models. \review{In Earth's magnetosphere, the plasma is dominated by its core magnetic field, thus the first feature we took into account is the magnetic field vector $\Vec{B}$. Additionally, in each magnetospheric region the plasma has a rather typical distribution function. However, the full particle distribution function constitutes an enormous amount of data when dealing with several years of data (as done here), in particular with the MMS mission (several TB of data). \review{Furthermore, important differences exist between different heliospheric missions in the specificities of the distribution functions, e.g. energy and angular ranges as well as energy and angular resolutions.} Therefore, we chose not to consider those products and focused on moments only (density, bulk velocity and temperature), to keep our model lightweight and as general as possible (i.e., applicable to different heliospheric missions such as Cluster). For this latter reason, we also decided not to use the spacecraft location as a feature, because different missions cover different locations in near-Earth space (e.g. Cluster has a polar orbit while MMS has an equatorial orbit; apogees/perigees are different, etc).}

We start with loading the raw MMS magnetic field and plasma moments data from FGM and FPI instruments for the 2016-2019 period using the aidapy package\footnote{\url{https://gitlab.com/aidaspace/aidapy}}, and resample the data to the FPI cadence, i.e., 4.5s. The resulting dataset contains 12 variables: the magnetic field magnitude $B_{tot}$ and its components $B_x$, $B_y$, $B_z$, the ion density $N_i$, the bulk velocity magnitude $V_{i_{tot}}$ and its components $V_{i_x}$, $V_{i_y}$, $V_{i_z}$, the parallel and perpendicular temperatures $T_\parallel, T_\perp$ with regards to the ambient magnetic field and the total value $T_{tot}$. The dataset is matched with the labels points defined in the previous section. \review{If data gaps are present in the specified time interval, then the whole interval is rejected. This method yields a total dataset of size $1,331,133\times12$.}




The TSC model requires as input arrays of equal sizes with one label per array. Thus the dataset is grouped as time series corresponding to the labeled time intervals, \review{resulting in $6,928$ data blocks of various sizes. We decided to split these time series into equal chunks of 3 minutes, because this time length correspond roughly to the mean value of the data blocks (i.e., it minimizes the padding) and yields enough points (40) in each block given the time sampling resolution. To do so, we apply the following scheme: if the data block is shorter than 3 minutes, it is padded with the wrap of the vector along the axis (i.e., the first values are used to pad the end and the end values are used to pad the beginning);
and if the data block is longer than 3 minutes, then it is split into several time series of 3 minutes each (the last one being padded as described above if also shorter than 3 minutes).~\autoref{tab:occurences} gives the distribution of the occurrences for each class.}

This operation yields a dataset as an multidimensional array of size (34159, 40, \review{12}). Finally, to input this array to the TSC model (and to avoid temporal bias due to the spacecraft orbit), we shuffle and randomly split the time series into training (56.25\%), test (25\%) and validation (18.75\%) datasets. \review{These three classical categories are defined such as:
\begin{itemize}
    \item the training set is used to determined the weights of the model;
    \item the validation set is an out-of-sample set and allows to evaluate the error for data which has not been observed. In particular, the validation set is used for hyperparameter search and early stopping to avoid over-fitting ensuring a good generalization of the model; 
    \item the test set is also an out-of-sample set which has not been used during training. The data are only used after the training to assess the performance of the final model, after training and hyperparameter optimization.
\end{itemize}
}


\begin{table}[h]
	\centering
	\begin{tabular}{c|c|}
		\cline{2-2}
		
		& \multirow{1}{*}{\begin{tabular}[c]{@{}c@{}}Time series\end{tabular}}\\
 \cline{1-2}
		
		\multicolumn{1}{|c|}{SW}           &  2,130       \\ \hline
		\multicolumn{1}{|c|}{FS}            &  4,714       \\ \hline
		\multicolumn{1}{|c|}{BS}           &  2,328      \\ \hline
		\multicolumn{1}{|c|}{MSH}           &  4,274       \\ \hline
		\multicolumn{1}{|c|}{MP}           &  3,764       \\ \hline
		\multicolumn{1}{|c|}{BL}          &  4,544        \\ \hline
		\multicolumn{1}{|c|}{MSP}           &  4,658        \\ \hline
		\multicolumn{1}{|c|}{PS}           &  3,576        \\ \hline
		\multicolumn{1}{|c|}{PSBL}           &  2,703        \\ \hline
		\multicolumn{1}{|c|}{LOBE}           &  1,468        \\ \hline
		\multicolumn{1}{|c|}{Total}       &  34,159        \\ \hline
	\end{tabular}
	\caption{Number of occurrences for each class.}
	\label{tab:occurences}
\end{table}

\section{Classification methods with a fully convolutional neural network}
In this section, we overview the machine-learning algorithm we use: the fully convolutional neural network (FCN). This algorithm is assessed with several evaluation metrics. FCN \review{works with time-dependant inputs (time series) in order to learn dynamical features.} The main purpose of classifying time series and not only the instantaneous values is to learn automatically different features across the time dimension. The events can be characterized by dynamical insights at different scales learned by the model. We present a second model, called Multilayer perceptron (MLP), in the supplemental materials. It considers \review{the input of the \review{$12$} variables as time independent}. The main advantages of this method are its simplicity to prepare the machine learning pipeline and the possibility to work with any temporal resolution. However, the classification performance is much lower.

The FCN belongs to the category of artificial neural networks. They are usually characterized by their architecture: number of layers, type of connection (feedforward, feedback, convolution, etc.), and the activation functions. A layer is defined as a set of neurons which are not connected to each other. We selected neural network as they are able to learn non-linear models and can handle large datasets compared to kernel methods.

\subsection{Model presentation}
Time series classification (TSC) is an active field of research and hundreds of different algorithms have been developed in recent years~\citep{bagnall2017great}. Classical TSC usually uses new features spaces generated from the time series. The best performing models can combine dozen of different classifiers and feature transformations, leading to significant complexity and an important computational cost~\citep{lucas2019proximity}.

Even if deep learning has seen very successful applications in the last decades, only a few examples of algorithms for TSC exist. This lack of overview has been filled recently~\citep{fawaz2019deep} and several models show very promising results. In particular, we are convinced that convolution architectures can build very accurate classifiers. The convolutional layers use convolution in place of general matrix multiplication.

Local groups of values are often highly correlated in time series and they form distinctive local motifs that are easily detected. \review{Convolutional layers are designed to identify these patterns invariant to location: if a motif specific to a label appears in one part of the time series, it could also appear anywhere. Hence a convolutional layer sharing the same weights at different locations can detect temporal patterns in different parts of the time series~\citep{lecun2015deep}.} Moreover, several deep learning frameworks, such as Tensorflow~\citep{abadi2016tensorflow} or Pytorch~\citep{paszke2019pytorch}, are nowadays freely available. They provide differentiable programming to build very efficient neural networks. Thus, we consider deep learning for TSC for these two reasons: the availability of high-quality frameworks and the relevance of convolutional architectures.

The Fully Convolutional Network (FCN)~\citep{wang2017time} is a competitive deep-learning architecture for TSC yielding good results on large benchmarks~\citep{fawaz2019deep}. The network architecture is relatively simple and is comprised of a sequence of thee time convolution blocks followed by a global average pooling block\review{~\citep{lin2013network}}. Each time convolution block is divided into a convolution layer, a batch normalization layer, and a Rectified Linear Unit (relu)~\cite{nair2010rectified} activation function. The convolutional layer is used to extract temporal features from the inputs by performing a convolution between the input signal and several filters. Each convolutional block has different filter lengths in order to analyze several time scales. The extracted features of the last convolutional block are used as inputs for the global average pooling (GAP) module to output the classification result. This last block is composed of the global average pooling layer and the Softmax layer. A more complete description of the FCN architecture can be found in~\citet{wang2017time, fawaz2019deep}.

\begin{figure}
    \centering
    \includegraphics[width=0.5\textwidth]{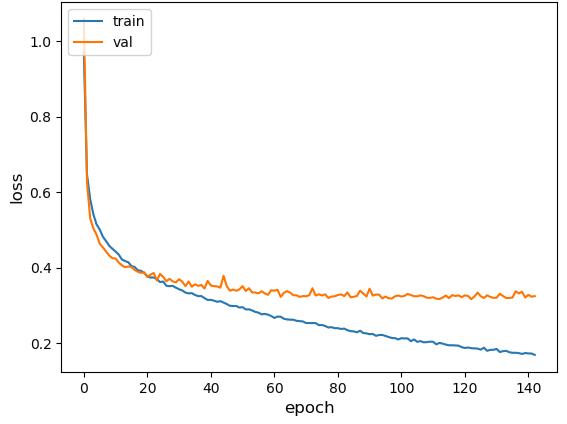}
    \caption{\review{Learning curve history comparing training and validation loss.}}
    \label{fig:learning_fcn}
\end{figure}

As regards the numerical parameter, we performed a \review{bayesian optimisation} to select the hyperparameters using the validation set for four parameters:
\review{
\begin{itemize}
    \item the optimizer: adam or rmsprop
    \item the learning rate: between $10^{-6}$ and $10^{-3}$;
    \item the batch size: 32, 64, or 92;
    \item the number of filters: the baseline of the FCN paper, half the baseline and twice the baseline. The baseline given by the FCN paper~\citep{wang2017time} is the following: the first convolutional block has 128 filters with a filter length equal to 8, the second convolution extracts features with 256 filters of size 5 and the final convolutional layer is defined by 128 filters, each one with a length equal to 3.
\end{itemize}
The final FCN is trained for $1,000$ epochs with the categorical cross entropy as loss function. The result of the bayesian opimization provides the different hyperparameters. The Adam optimizer is used
with an initial value of the learning rate set to $0.00002$. An adaptive learning rate decrease is selected (with a minimum learning rate of $10^-{6}$), reducing the learning rate when the validation accuracy has stopped improving for 50 epochs. A batch size of $92$ is used with twice the baseline configuration for the number of filters. The final FCN has $1,079,000$ free parameters. The method of early termination is adopted to avoid over-fitting and stops the learning process when the validation loss did not decrease for 20 epochs. All the time series are standardized before the training process. The learning curve is presented in~\autoref{fig:learning_fcn}. The learning process stopped after $140$ epochs thanks to the early termination strategy.}

\subsection{Performance evaluation}
Assessing accurately the quality of a classification model raises significant challenges. A single and unified performance metric cannot effectively evaluate a multi-class classifier from all perspectives, such as class-balance or the number of different outcomes. Therefore, several metrics generalized from binary classification are considered.

The confusion matrix is a basic but useful tool to visualize the distribution of the predicted classes. Each column of the matrix gives the instances of a true class while each row represents the instance of a predicted class. Several metrics from binary classification can be generalized to multiclass by considering the prediction of each class as a binary classification problem: the given class has been predicted (positive) or not (negative). Then, the total number of True Positive (TP), True Negative (TN), False Positive (FP), and False Negative (FN) can be computed and several metrics are defined: 
\begin{itemize}
	\item [-] \textbf{Accuracy} is the most intuitive metric and gives the percentage of correct predictions among all the predictions $N_p$:
	\begin{equation}
	accuracy = \frac{TP + TN}{N_p}.
	\end{equation}
	\item [-] \textbf{Precision}, also called positive predictive value, measures the proportion of positive values that are correctly identified as such among all predicted positives. In other words, it corresponds to the percentage of detections that are right. High precision is associated to a small amount of FP:
	\begin{equation}
	precision = \frac{TP}{TP+FP}.
	\end{equation}    
	\item [-] \textbf{Recall}, also called the true positive rate, measures the proportion of positives detected. A significant number of missed true values (FN) leads to low recall: 
	\begin{equation}
	recall = \frac{TP}{TP+FN}.
	\end{equation}
	\item [-] \textbf{F1-score} is the harmonic mean of precision and recall. Usually, high recall is detrimental to precision and vice versa. Therefore, the F1-score gives a trade-off between the two quantities and can help to compare two classifiers with different precision and recall values:
	\begin{equation}
	accuracy = 2 . \frac{precision . recall}{precision + recall}.
	\end{equation}
\end{itemize}
All these metrics are computed for each class. Therefore, averaged quantities can be also defined to provide a more general view of the classifier. A macro-average gives the unweighted mean of the metric without taking imbalance into account and a micro-average computes the average weighted by the number of true instances for each class. The micro-average will account for label imbalance.

Finally, a receiver operating characteristic (ROC), also called ROC curve, is a graphical tool illustrating the performance of a binary classifier system when the discrimination threshold is varied. Indeed, each class prediction is associated to a probability. Usually, the class predicted by the classifier is chosen as the class with the highest probability. Instead, a threshold value can be used to decide if the class is effectively predicted by the classifier. The ROC curve plots the fraction of true positives out of all the predicted positives (True Positive Rate) versus the fraction of false positives out of all the predicted negatives (False Positive rate) at various threshold values. For instance, all classes are predicted with a zero threshold (False Positive Rate and True Positive Rate of 1.0) while there are no correct predicted class and no false positive with a unit threshold (False Positive Rate and True positive rate of 0.0). A perfect classifier is located on the left corner of the curve with a true positive rate of 1.0 and a false positive rate of 0.0. A ROC curve can be drawn for each class. In case of a model with no discrimination capacity to distinguish between positive class and negative class, the ROC curse is a straight line of slope 1. The area under the ROC curve (AUC) summarized in one number if the model is capable of distinguishing between classes.

\section{Results}

\subsection{Classification of MMS data}
\review{We use here the test set (the out-of-sample set that is not involved in the training process) to evaluate the performance of the model. It is obtained by shuffling 25\% of the data as described in section $2.3$.}

\begin{table}[h]
	\centering
	\begin{tabular}{c|c|c|c|c|}
		\cline{2-5}
		& Precision & Recall & f1-score & AUC     \\ \hline
		\multicolumn{1}{|c|}{SW}            & 0.88      & 0.92   & 0.90     & 1.00 \\ \hline
		\multicolumn{1}{|c|}{FS}            & 0.93      & 0.95   & 0.94     & 1.00 \\ \hline
		\multicolumn{1}{|c|}{BS}            & 0.85      & 0.80   & 0.83     & 0.99 \\ \hline
		\multicolumn{1}{|c|}{MSH}           & 0.91      & 0.91   & 0.91     & 0.99 \\ \hline
		\multicolumn{1}{|c|}{MP}            & 0.84      & 0.84   & 0.84     & 0.99 \\ \hline
		\multicolumn{1}{|c|}{BL}            & 0.84      & 0.84   & 0.84     & 0.98 \\ \hline
		\multicolumn{1}{|c|}{MSP}           & 0.96      & 0.95   & 0.96     & 1.00 \\ \hline
		\multicolumn{1}{|c|}{PS}            & 0.89      & 0.89   & 0.89     & 0.99 \\ \hline
		\multicolumn{1}{|c|}{PSBL}          & 0.86      & 0.87   & 0.87     & 0.99 \\ \hline 
		\multicolumn{1}{|c|}{LOBE}          & 0.95      & 0.95   & 0.95     & 1.00 \\ \hline \hline
		\multicolumn{1}{|c|}{Macro average} & 0.89      & 0.89   & 0.89     & 0.98 \\ \hline
		\multicolumn{1}{|c|}{Micro average} & 0.89      & 0.89   & 0.89     & 0.99 \\ \hline
		\multicolumn{1}{|c|}{Accuracy} & \multicolumn{4}{|c|}{0.89}     \\ \hline
	\end{tabular}
	\caption{Classification report for FCN \review{computed on the test set}.}
	\label{tab:report_fcn}
\end{table}

\review{\autoref{tab:report_fcn} summarizes the precision, recall, f1-score, and AUC evaluated by the FCN
for the 10 different classes. In term of absolute values, the FCN shows great performance metrics. The AUC value is above $0.98$ for all the classes. It means that a significant probability is always associated to the correct region by the classifier, even it is not the highest probability. For instance, the classifier could give a probability of 45\% to the SW and 35\% to the FS for a test sample belonging to the FS. Even if the prediction is wrong, the classifier still
has detected some features which can belong to the correct class.}

\begin{figure}
	\centering
	\includegraphics[width=0.5\textwidth]{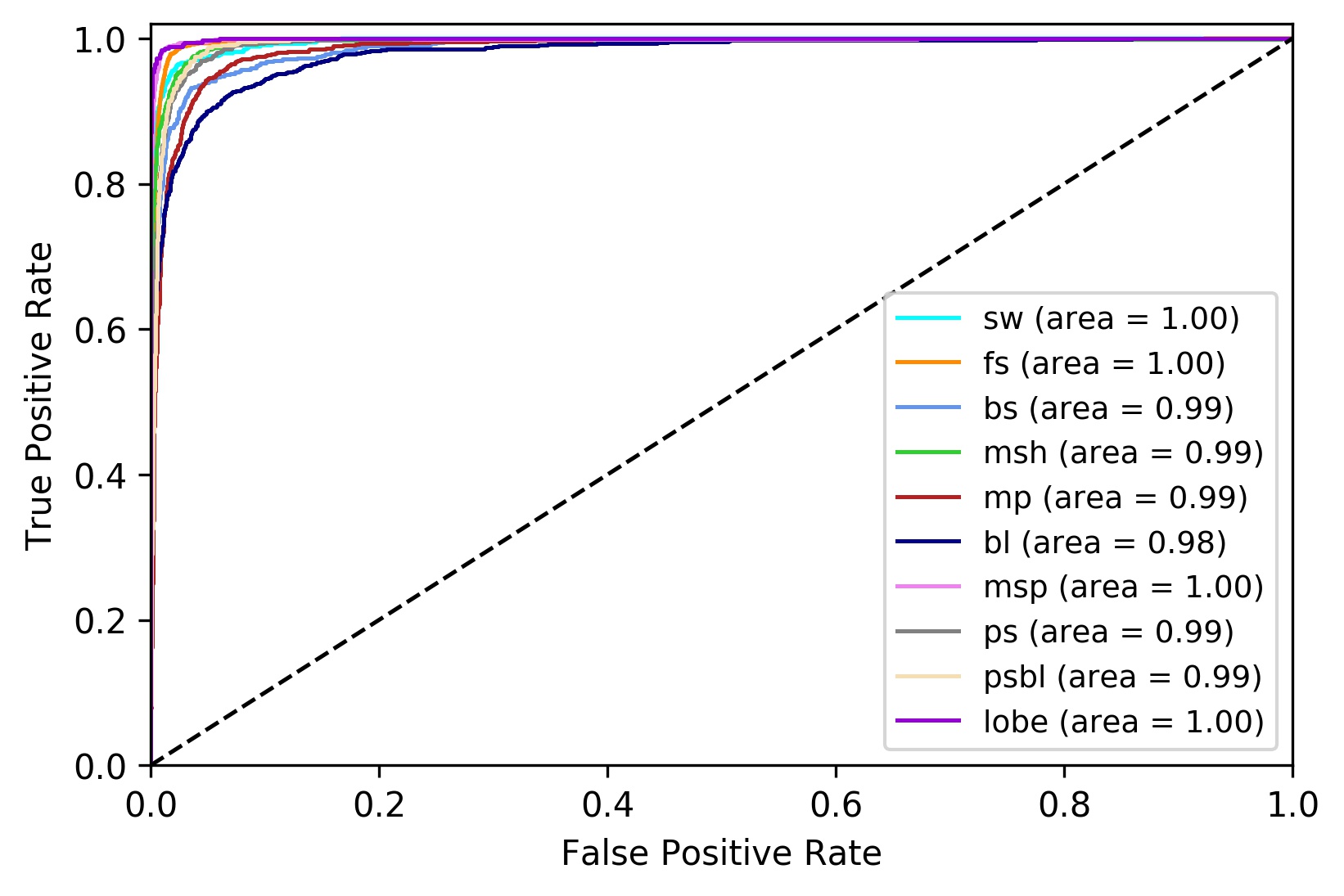}
	\caption{Receiver operating characteristic curve computed on the test set.}
    \label{fig:roc_fcn}
\end{figure}

\review{Two different groups of regions are identified in~\autoref{tab:report_fcn}. The first group consisting of the FS, the MSP, and the LOBE is very well predicted with a respective f1-score above $0.94$ and a AUC of $1.00$. Such regions show usually very specific patterns, explaining why the classification metrics are very high. The second group is formed by the BS, the MP, the BL, the PS, and the PSBL, with smaller metric values, such as a f-1 score between $0.82$ and $0.88$ and a AUC between $0.98$ and $0.99$. 
~\autoref{fig:roc_fcn} shows the different ROC curves for the FCN. BL and BS are the lowest curves. They have a much higher false positive rate when they reached a true positive rate of $1.0$ compared to the other classes. It means that, for some very specifics predictions, the classifier associates a very low probability to the correct class. The BL and the BS are thin regions marking the boundary between larger regions. Therefore, several time intervals labeled by SITLs may overlap two or more nearby regions. It could explain the lower metric values for these two regions.}

\review{From the supplemental material, the second model (MLP) has important also problems to classify regions with strong variations, such as the BS with a f-1 score below $0.5$.} Thus, it seems that convolutional methods help to improve the accuracy of the classifier for the most challenging regions and it validates the choice of such algorithms designed to extract important temporal features and variations.
 
\review{~\autoref{tab:report_fcn} gives details on the FCN predictions with a confusion matrix. Very typical errors can be observed for nearby regions, such as: SW and FS; PS, BL, and PSBL; PSBL and LOBE; MP, MSH and BL. These errors can be explained by physical arguments. The plasma properties can be similar for several nearby regions.
For instance, we may expect that MP shares common properties with MSH and
MSP as the MP acts as a boundary between the two other regions. The SW
region can also be very similar to the FS as the latter comes from the reflection
of the SW on the BS, as stated in the introduction. Therefore, it may be concluded that the FCN model really learned the typical patterns associated to each region as the main errors are identified for regions with similar physics.}

\begin{table}[]
	\centering
	\begin{tabular}{cl|c|c|c|c|c|c|c|c|c|c|}
		\cline{3-12}
		&     & \multicolumn{10}{c|}{True}                       \\ \cline{3-12} 
		&     & SW    & FS & BS  & MSH  & MP  & BL  & MSP & PS  & PSBL & LOBE\\ \hline
		\multicolumn{1}{|c|}{\multirow{10}{*}{\begin{sideways}Prediction\end{sideways}}}
		& SW  &\cellcolor[HTML]{C0C0C0} 493 & 53  & 8  & 0   & 2   & 1   & 0   & 0 & 0 & 0\\ \cline{2-12} 
		\multicolumn{1}{|c|}{}                            & FS & 42    & \cellcolor[HTML]{C0C0C0}1,140 & 40   & 0  & 0   & 0  & 0   & 0  & 0 & 0\\ \cline{2-12} 
		\multicolumn{1}{|c|}{}                            & BS  & 2    & 12   & \cellcolor[HTML]{C0C0C0}450 & 54   & 8   & 1   & 0   & 0  & 0 & 0\\ \cline{2-12} 
		\multicolumn{1}{|c|}{}                            & MSH  & 0     & 1  & 52   & \cellcolor[HTML]{C0C0C0}982 & 41   & 5  & 0   & 0  & 0 & 0 \\ \cline{2-12} 
		\multicolumn{1}{|c|}{}                            & MP  & 0     & 0   & 10   & 37  & \cellcolor[HTML]{C0C0C0}801 & 92  & 5   & 3   & 2 & 1\\ \cline{2-12} 
		\multicolumn{1}{|c|}{}                            & BL  & 0     & 0   & 1   & 4  & 78  & \cellcolor[HTML]{C0C0C0}933 & 49   & 39   & 10 & 0\\ \cline{2-12} 
		\multicolumn{1}{|c|}{}                            & MSP & 0     & 0   & 0   & 0   & 10   & 32   & \cellcolor[HTML]{C0C0C0}1,101 & 0  & 0 & 0\\ \cline{2-12} 
		\multicolumn{1}{|c|}{}                            & PS & 0     & 0   & 0   & 0   & 5   & 38   & 5 & \cellcolor[HTML]{C0C0C0}803  & 54 & 1\\ \cline{2-12} 
		\multicolumn{1}{|c|}{}                            & PSBL & 0     & 0   & 0   & 0   & 7   & 13   & 0 & 59  & \cellcolor[HTML]{C0C0C0}572 & 15\\ \cline{2-12} 
		\multicolumn{1}{|c|}{}                            & LOBE  & 0    & 0   & 0  & 0   & 0   & 0   & 0  & 1 & 16 & \cellcolor[HTML]{C0C0C0}354\\ \hline
	\end{tabular}
	\caption{Confusion Matrix for FCN \review{computed on the test set}.}
	\label{tab:confusion_fcn}
\end{table}

\subsection{Examples of MMS classification}
In this section we show some examples of time series observed by the MMS spacecraft as it spanned the different near-Earth regions, along with the labels of the classifications made by \review{the FCN model}.
The labels of the different regions are sorted roughly according to their distance to the Sun, and summarized in \autoref{tab:region_labels}.

\begin{figure}[]
	\centering
	\includegraphics[width=\textwidth]{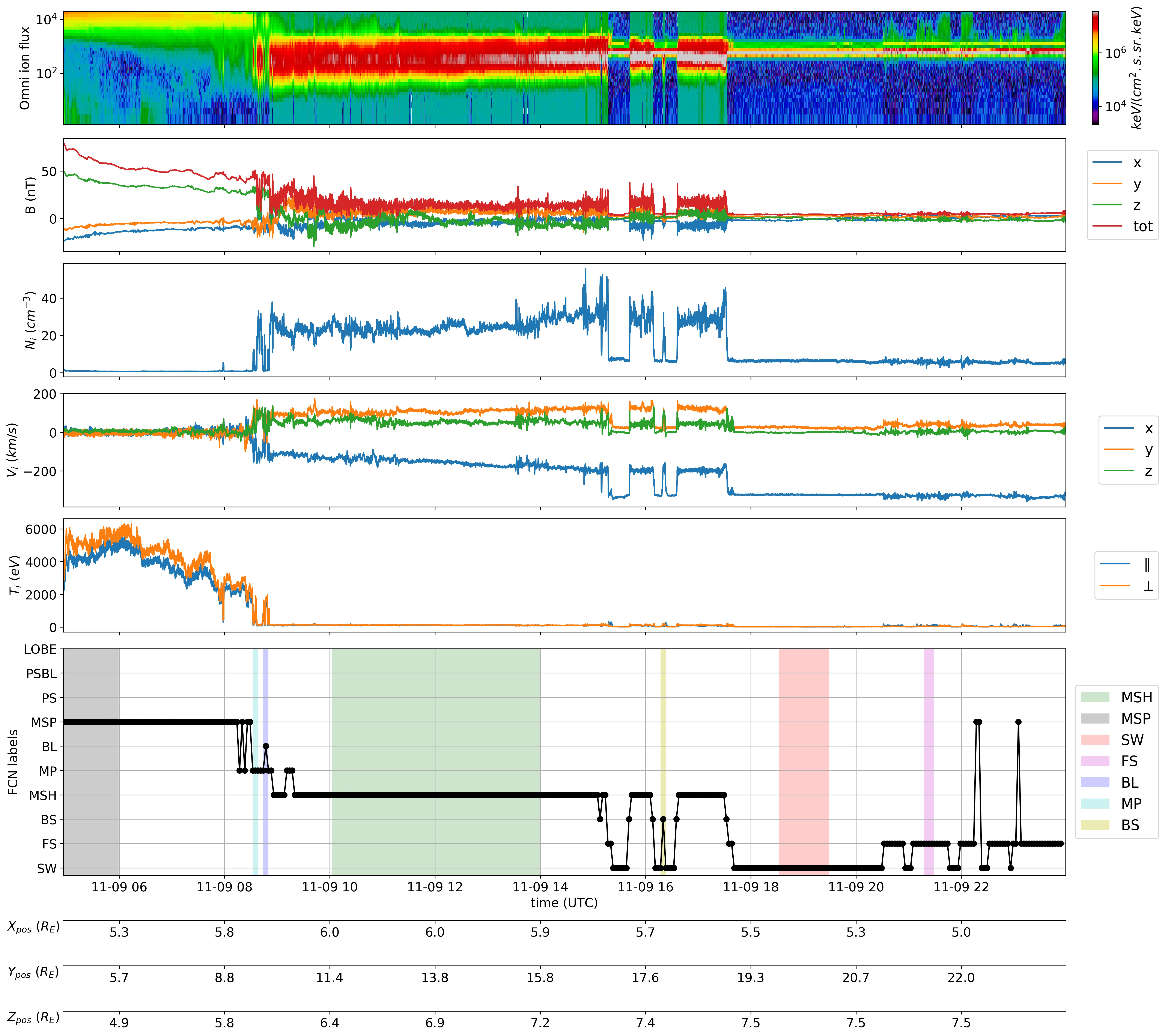}
	\caption{Example of MMS dayside classification. \review{Labels, not present in the initial data set, have been manually added to compare the FCN predictions with a ground truth. They are represented by colored regions on the FCN labels plot.}}
	\label{fig:mms_dayside}
\end{figure}

\autoref{fig:mms_dayside} shows the MMS1 probe observations for the whole day of 2019-11-09, during which it flew through the dayside region. From top to bottom, the figure panels show the energy spectrum of the omnidirectional ion flux, the three components of the magnetic field ($B_x, B_y, B_z$) in GSE coordinates and its magnitude $B_{tot}$, the ion number density, the three components of the ion bulk velocity ($V_x, V_y, V_z$) in GSE coordinates, the ion temperatures parallel and perpendicular ($T_\parallel, T_\perp$) to the ambient magnetic field, and finally the labels given by the FCN model (FCN labels). The time in UT format and the spacecraft position ($X_{pos}, Y_{pos}, Z_{pos}$ in $R_E$ and GSE coordinates) are displayed at the bottom of the X-axis.

In this example, the FCN model classifies well the regions, as its labels follow to a great extend the MMS1 observations: first the spacecraft is located in the magnetosphere, then it crosses the magnetopause to enter the magnetosheath, afterwards it flies through a series of bow shocks and goes back and forth to the solar wind, until it finally enters the solar wind. The model is even able to precisely detect the small-scale dynamics of the boundary regions such as partial magnetopause, boundary layer and bow shock crossings, and the come and go between the solar wind and the ion foreshock. Only three points seem to be clearly misclassified (as MSP) in the ion foreshock. However, these points can be easily eyeballed as outliers or discriminated by their low quality flag.

\begin{figure}[]
	\centering
	\includegraphics[width=\textwidth]{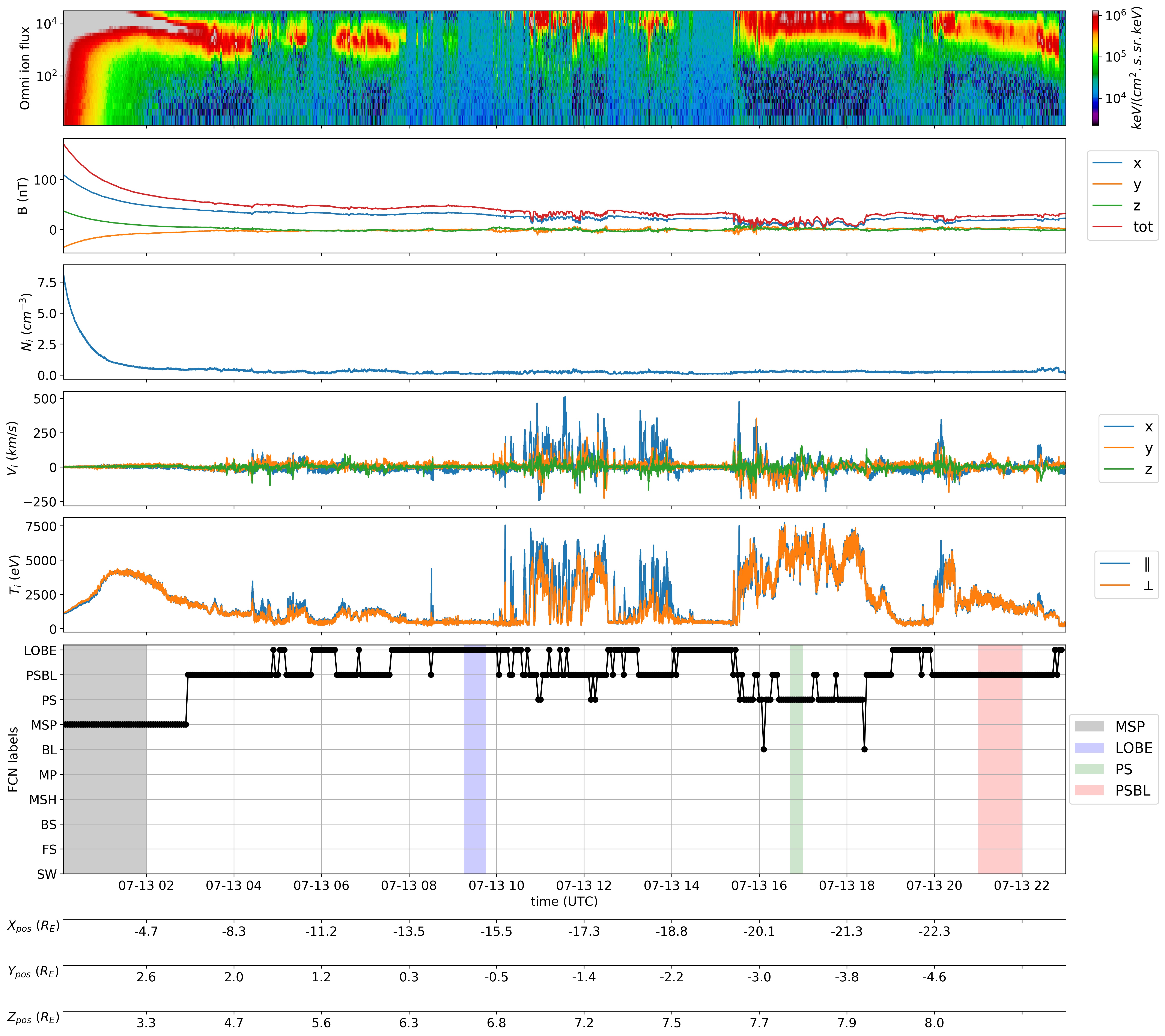}
	\caption{Example of MMS nightside classification. \review{Labels, not present in the initial data set, have been manually added to compare the FCN predictions with a ground truth. They are represented by colored regions on the FCN labels plot.}}
	\label{fig:mms_nightside}
\end{figure}

\autoref{fig:mms_nightside} shows another example of the FCN classification method described above, but this time when MMS spacecraft flew through the nightside region. The MMS1 observations take place during the whole day of 2019-07-13, and the format is the same as in \autoref{fig:mms_dayside}. In this example again, the FCN model is able to label properly the regions spanned by MMS1: at the very beginning MMS1 exits the magnetosphere to the plasma sheet boundary layer, then goes in and out the lobe regions before reaching the plasma sheet. Again, the FCN model is also able to pick up the small-scale dynamics such as short plasma sheet crossings.

The two examples shown in this section therefore support the good metrics obtained in the previous section, showing concretely that the FCN model is suited to properly label most of the MMS data with the regions spanned.

\subsection{Locating important feature in time series with class activation map}
\review{Class Activation Map (CAM) is a technique developed by~\citet{zhou2016learning} to get the discriminative regions used by a CNN to identify a specific class in the input data. CAM has been applied to TSC for one-dimensional case in~\citep{wang2017time} for the first time. The objective in our case is to highlight the subsequences (i.e., sections) in each time series which are relevant to its class. Thus, this approach \review{allows} to explain the decision taken by the classifier. The mathematical description of the method can be found in~\cite{zhou2016learning, wang2017time, fawaz2019deep}. This method works only for models with the global average pooling as last layer\review{~\citep{lin2013network}}. CAM is represented by a univariate time series (with the same size than the FCN input) where the value gives the importance of the signal to classification. Low values mean the subsequence does not contribute to the decision of the classifier while high values mean the section contributes significantly.}


\begin{figure}[]
	\begin{minipage}{0.42\textwidth}
		\centering
		\includegraphics[width=\textwidth]{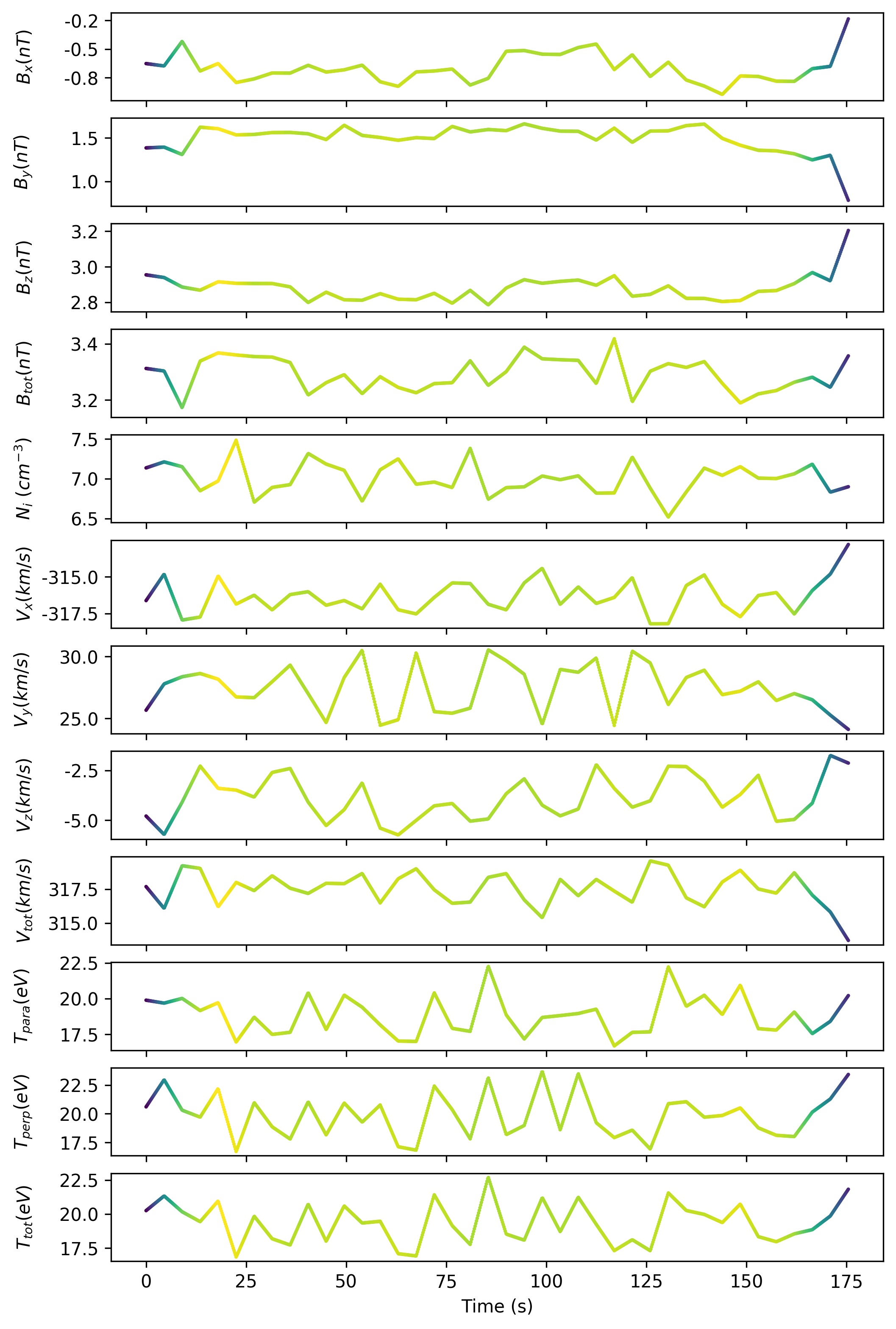}
		\subcaption{Solar wind.}
		\label{fig:cam_sw}
	\end{minipage}\hfill
	\begin{minipage}{0.42\textwidth}
		\centering
		\includegraphics[width=\textwidth]{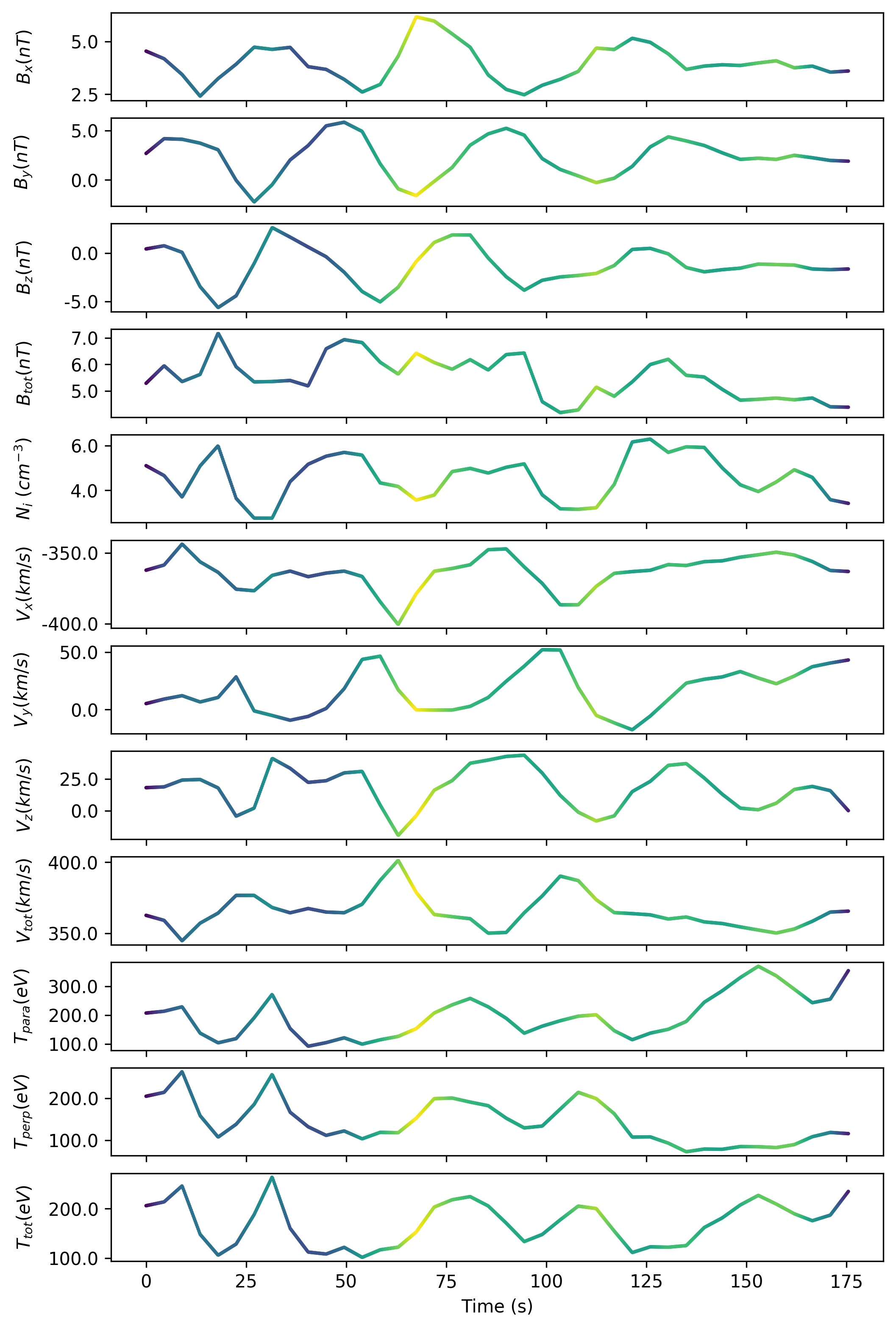}
		\subcaption{Foreshock.}
		\label{fig:cam_fs}
	\end{minipage}
		\begin{minipage}{0.15\textwidth}
		\centering
		\includegraphics[width=0.4\textwidth]{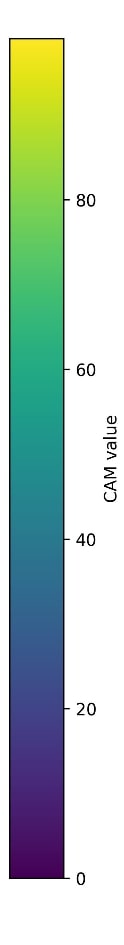}
	\end{minipage}
	\caption{The Class Activation Map highlighting the important sections of the time series contributing most to the classification. \review{Low values mean the time step of the time series does not contribute to the decision of the classifier while high values mean the time step contributes significantly.}}
	\label{fig:cam}
\end{figure}
~\autoref{fig:cam} shows CAM examples for two different classes: the solar wind and the foreshock. For each class, a well-predicted time series is drawn and colored by the value of the CAM, illustrating the results from MMS data given in section $4.1$. \review{As regards the SW example, almost all the input signal contributes to the classification, as stated by the uniform CAM values above 0.8 between 20 and 160 seconds. It means that the values of input features are much more important than their variations or than specific patterns. On the other hand, the classifier detects the FS mainly thanks to a specific pattern around 65 seconds associated to extrema for $B_x$, $B_y$, $N_i$, and $V_y$ and significant slopes for the other features. Only a few specific time intervals are involved in the decision of the classifier for FS. The CAM analysis strengthens the conclusion that temporal and dynamical analysis are fundamental to classify nearby regions with fluctuating behaviors. For instance, the MLP results given in the supplemental materials shows that a significant number of SW region are misclassified as FS as the model works only on instantaneous quantities.}

\subsection{Extension to Cluster mission}
In this section, we put to test the adaptability of the FCN model
to different magnetospheric data, namely data from the Cluster mission. This mission has been chosen as the data is pretty close to MMS data (notably the data sampling), and a 2-year period of labeled data is available (see~\citet{nguyen2019}) to audit the classification resulting from the model.

\begin{figure}[]
	\centering
	\includegraphics[width=\textwidth]{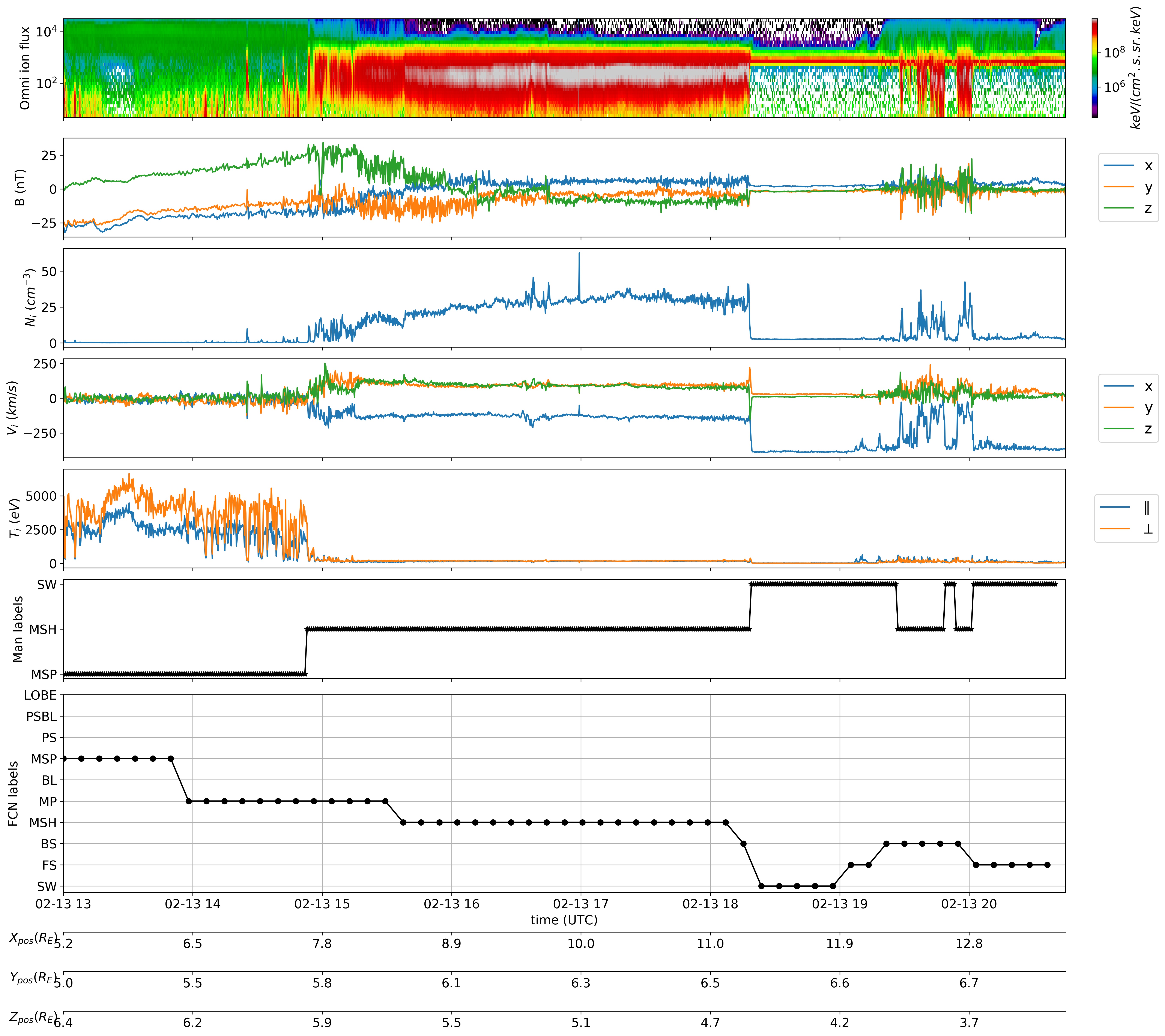}
	\caption{Example of Cluster dayside classification.}
	\label{fig:cluster_dayside}
\end{figure}

First, we qualitatively assess the quality of the different classification methods, by comparing them with C1 probe observations. We randomly select a day during which the C1 probe flew through the dayside region. This example, from the day of 2005-02-13, is shown in \autoref{fig:cluster_dayside} in the same format as \autoref{fig:mms_dayside} and \autoref{fig:mms_nightside}, except that we include the manual labels from~\citet{nguyen2019} (named "Man labels") in the penultimate bottom panel. 

The "Man labels" are in good agreement with the observations of C1, as the probe is first located in the magnetosphere, then flies through the magnetosheath and finally reaches the solar wind. The "FCN labels" show also good agreement with C1 observations: the probe is first located in the magnetosphere and then crosses the magnetopause a few times before entering the magnetosheath, then it crosses the bow shock to enter the solar wind before crossing several times again the bow shock and finally spanning the ion foreshock. As in the examples shown for the MMS mission, the FCN model is able to pick up the small-scale dynamics of the near-Earth frontiers: the magnetopause, the bow shock and the foreshock. As a result, the model is able to outperform the manual labels by identifying in more detail the magnetospheric regions, in particular the small-scale dynamics of the transition regions.

Then, we take advantage of the 2-year period of Cluster labeled data from~\citet{nguyen2019} to quantify the quality of the FCN model classification, by investigating the performance metrics. We select the points labeled as 'MSP', 'MSH' and 'SW' by the FCN model using Cluster data from the 2005-2006 period, and compare them with the labels for these 3 regions obtained from Nguyen. \review{Using this method, we obtain an accuracy classification score of 0.97 over a total of $10,929$ common labeled points. the classification report can be found in \autoref{tab:report_cluster}}. 
\begin{table}[h]
	\centering
	\begin{tabular}{c|c|c|c|}
		\cline{2-4}
		& Precision & Recall & f1-score     \\ \hline
		\multicolumn{1}{|c|}{MSP}            & 0.84      & 0.97   & 0.90     \\ \hline
		\multicolumn{1}{|c|}{MSH}            & 1.00      & 0.97   & 0.98     \\ \hline
		\multicolumn{1}{|c|}{SW}          & 0.95      & 0.99   & 0.97     \\ \hline \hline
		\multicolumn{1}{|c|}{Macro average} & 0.93      & 0.97   & 0.95     \\ \hline
		\multicolumn{1}{|c|}{Micro average} & 0.98      & 0.97   & 0.97     \\ \hline
		\multicolumn{1}{|c|}{Accuracy} & \multicolumn{3}{|c|}{0.97}     \\ \hline
	\end{tabular}
	\caption{Classification report for FCN \review{computed on the Cluster dataset}.}
	\label{tab:report_cluster}
\end{table}
We note here that this score is not an exact quantization of the model's accuracy on Cluster data\review{, and the classification score is probably overestimated because only 3 classes (i.e., magnetospheric regions) are considered here instead of the 10 included in the FCN model.} Therefore, a dataset of Cluster data labeled with these 10 near-Earth regions is required to get an absolute quantification of its accuracy regarding the Cluster mission. However, it shows globally that our FCN model classification is a reliable method for the labeling of the Cluster mission data and potentially other heliophysics missions.


\subsection{Ressources}
The most time-consuming task in the present work has been the data preparation and the parsing of the SITL reports.
Training a FCN model needs about $200$ CPU hours on a workstation featured with an Intel Xeon E5-2670 v3 (12 cores at  2.30GHz). These values must be multiplied by the number of different trials needed to optimize the hyperparameters (about 40 in that case).

\section{Discussion and conclusions}
Using deep learning algorithms, namely a fully convolutional neural network (FCN), we built an automatic detection of \review{10} near-Earth regions: \review{the solar wind, the ion foreshock, the bow shock, the magnetosheath, the magnetopause, the boundary layer, the magnetosphere, the plasma sheet, the plasma sheet boundary layer and the lobes.}

Using more than 3 years of labeled (SITL and additional human-labeled) MMS mission data, we showed that this method are reliable to classify near-Earth regions. \review{The FCN method(1,079,000 free parameters) has been very effective in taking into account the dynamical features of the most challenging plasma regions (important data variability due to the fluctuations of the plasma), in particular the bow shock and the boundary layer. The high accuracy of the FCN model also highlights the quality of the labeled data set generated from SITL reports.} We demonstrate the good accuracy of the classification predictions on the test and validation datasets, but also on unlabeled data from the MMS mission. We also show the adaptability of the trained FCN model by applying the classification to the Cluster ESA mission data. The predictions showed good accuracy and enhanced dynamics in comparison with previous human-labeled dataset.

These results show on one hand that the model can be applied to the whole MMS dataset, which would be of great interest to map the different magnetospheric boundaries and build empirical models of the properties and dynamics of the plasma in the near-Earth space. On the other hand, these results show that the model could be applied on different space missions orbiting around Earth (such as Cluster or THEMIS) or other planets to automatically label available data. However, \review{we recommend} to build a small labeled dataset specific to each mission with limited retraining of the last layer. This process called transfer learning could help improving the generalization of our labeled dataset to other missions such as the JUICE mission, which will be launched in 2022 and for which selective downlink strategies are being discussed. 

As a matter of fact, the model could be used as a support to the scientific experts in charge of the data selection process (such as the SITL system), providing classification of the regions that would contribute to ease and speed up this time-consuming and tedious task. \review{Such lightweight and easily-adaptable algorithms could also be important for the so-called SOC (Science Operation Center) of current and future spacecraft missions, both on ground and onboard. For the latter case, these algorithms could be implemented within onboard spacecraft digital boards to automatically select regions and events of scientific interest, much reducing the complexity and the cost of science operations.}

The intended integration of this model into the aidapy package\footnote{\url{https://gitlab.com/aidaspace/aidapy}}, which allows to automatically load selected data from open-access databases, will be of particular interest by providing plasma data accompanied with region labels and quality flags on the fly for use case and statistical studies. The integration of such lightweight algorithms can be also generalized to other programs.

In terms of modeling, we expect to improve the FCN results by generalizing the use of hyperparameter optimization for a higher number of hyperparameters and on larger ranges. We also plan to investigate the use of convolution methods directly on the omnidirectional ion energy fluxes.

Finally, the classification of near-Earth regions represents only a first step for the time series classification in heliophysics. It paves the way to more ambitious models, such as the identification and the classification of space plasma processes (e.g., magnetic reconnection and structures, waves and turbulence or plasma jets), as well as their combinations. However, this topic is beyond the scope of the present study and is left for a forthcoming study.

\section*{Conflict of Interest Statement}
The authors declare that the research was conducted in the absence of any commercial or financial relationships that could be construed as a potential conflict of interest.

\section*{Author Contributions}
\review{HB and RD built the data set, the machine learning models and wrote the manuscript. HB, RD, AR and OL gathered information from SITL reports and provided information for their usage for machine learning. HB, AR, OL and GL provided physical interpretation of the results. RD and JA provided insights into the use of the different machine learning techniques. AR and GL supervised the work. All authors contributed to manuscript revision, read and approved the submitted version.}

\section*{Funding}
This paper has received funding from the European Unions Horizon 2020 research and innovation programme
under grant agreement No 776262 (AIDA, \url{www.aida-space.eu}).

\section*{Acknowledgments}
Data were retrieved using HELIOPY v0.5.3\citep{stansby2020} and processed using aidapy\footnote{\url{https://gitlab.com/aidaspace/aidapy}}, pandas~\cite{mckinney2010}, and scikit-learn~\citep{pedregosa2011scikit}. Figures were produced using matplotlib~\citep{hunter2007matplotlib}. The model has been built with Keras from Tensorflow 2.2~\citep{abadi2016tensorflow} and we modified scripts provided in the dl-4-ts repository~\footnote{\url{https://github.com/hfawaz/dl-4-tsc}}. \review{The hyperparameter optimization is performed with the Python library optuna~\citep{akiba2019optuna}. The authors would like to acknowledge Gautier Nguyen (LPP) for providing classifications on Cluster data, Nicolas Aunai (LPP) and Alexis Jeandet (LPP) for the helpful discussion and support.}


\section*{Data and Model Availability Statement}
The models used for this study, the post-processing tools, \review{and the link to the data} can be found in the Jupyter notebook classification\_results.ipynb in the AIDA repository~\footnote{\url{https://gitlab.com/aidaspace/notebooks_aida/-/tree/master/04_sitl_classification_region}}.


\bibliographystyle{frontiersinHLTH&FPHY} 
\bibliography{references}


\end{document}